\documentclass[twocolumn]{aastex6}

\newcommand{\ergs}{erg~s$^{-1}$}

\newcommand{\cxo}{\textit{Chandra~}}

\begin{document}
\title{X-ray and Ultraviolet Properties of AGN in Nearby Dwarf Galaxies}

\author{Vivienne F. Baldassare\altaffilmark{1}, Amy E. Reines\altaffilmark{2,3}, Elena Gallo\altaffilmark{1}, Jenny E. Greene\altaffilmark{4}}

\altaffiltext{1}{Department of Astronomy, University of Michigan, Ann Arbor, MI 48109}
\altaffiltext{2}{National Optical Astronomy Observatory, 950 N Cherry Ave, Tucson, AZ 85719}
\altaffiltext{3}{Hubble Fellow}
\altaffiltext{4}{Department of Astrophysical Sciences, Princeton University, Princeton, NJ 08544}


\begin{abstract}

We present new \textit{Chandra X-ray Observatory} and \textit{Hubble Space Telescope} observations of eight optically selected broad-line AGN candidates in nearby dwarf galaxies ($z<0.055$). Including archival \textit{Chandra} observations of three additional sources, our sample contains all ten galaxies from Reines et al. (2013) with both broad H$\alpha$ emission and narrow-line AGN ratios (6 AGNs, 4 Composites), as well as one low-metallicity dwarf galaxy with broad H$\alpha$ and narrow-line ratios characteristic of star formation. All eleven galaxies are detected in X-rays. Nuclear X-ray luminosities range from $L_{0.5-7 \rm{keV}}\approx5\times10^{39}$ to $1\times10^{42}$ $\rm{erg}\rm{s^{-1}}$. In all cases except for the star forming galaxy, the nuclear X-ray luminosities are significantly higher than would be expected from X-ray binaries, providing strong confirmation that AGN and composite dwarf galaxies do indeed host actively accreting BHs. Using our estimated BH masses (which range from $\sim7\times10^{4}-1\times10^{6}~M_{\odot}$), we find inferred Eddington fractions ranging from $\sim0.1-50\%$, i.e. comparable to massive broad-line quasars at higher redshift. We use the \textit{HST} imaging to determine the ratio of ultraviolet to X-ray emission for these AGN, finding that they appear to be less X-ray luminous with respect to their UV emission than more massive quasars (i.e. $\alpha_{\rm OX}$ values an average of 0.36 lower than expected based on the relation between $\alpha_{\rm OX}$ and $2500{\rm \AA}$ luminosity). Finally, we discuss our results in the context of different accretion models onto nuclear BHs. 

\end{abstract}

\section{Introduction}

In the last few years, the number of active galactic nuclei (AGN) identified in dwarf galaxies (i.e., $M_{\ast}\lesssim3\times10^{9}M_{\odot}$) has grown from a handful of quintessential examples (e.g., NGC 4395; \citealt{1989ApJ...342L..11F}, and POX 52; \citealt{2004ApJ...607...90B}) to a body of several hundred candidates (see \citealt{2016arXiv160903562R} for a review). This has largely been possible thanks to large scale optical spectroscopic surveys (e.g., the Sloan Digital Sky Survey; SDSS), which have facilitated the search for AGN signatures in samples of tens of thousands of galaxies (see e.g. \citealt{2004ApJ...610..722G, 2007ApJ...670...92G, 2012ApJ...761...73D}), with the most recent studies concentrating on \textit{bona-fide} dwarf galaxies \citep{Reines:2013fj, 2014AJ....148..136M, 2015MNRAS.454.3722S}. In particular, the most successful searches for AGN in dwarf galaxies have used narrow emission line diagnostics (e.g., the BPT diagram; Baldwin, Phillips \& Terlevich 1989\nocite{1981PASP...93....5B}) to search for photo-ionized gas consistent with the presence of an AGN (see also \citealt{2003MNRAS.346.1055K, 2006MNRAS.372..961K} for commonly used diagnostics).  

For AGN exhibiting broad H$\alpha$ emission, assuming that the broad line region gas is virialized, it is possible to estimate the mass of the central BHs. Note that, for AGN in dwarf galaxies, this also relies on the assumption that the scaling relation between BH mass and host stellar velocity dispersion \citep{2000ApJ...539L...9F, 2000ApJ...539L..13G} holds in this mass regime \citep{2016arXiv160803893B}. The velocity of the broad line region gas is estimated from the width of the H$\alpha$ line, and the radius to the broad line region is estimated from the luminosity of the broad emission (\citealt{2000ApJ...533..631K, 2004ApJ...613..682P, 2005ApJ...630..122G,2009ApJ...705..199B, 2013ApJ...767..149B}). BH masses in dwarf AGN are typically $\sim10^{5}-10^{6}M_{\odot}$ solar masses (see e.g., \citealt{Reines:2013fj}, \citealt{2016arXiv160505731B}), with the lowest reported having a mass of just $\sim50,000 M_{\odot}$ \citep{2015ApJ...809L..14B}. 

Despite these recent advances in the identification of AGN in dwarf galaxies, the radiative properties of this population of AGN as a whole are largely unconstrained. Much work has been done exploring the X-ray properties of $\sim10^{6}M_{\odot}$ optically selected AGN from the Greene \& Ho samples \citep{:kx, 2012ApJ...761...73D, 2016ApJ...825..139P}, but these host galaxies tend to be more massive than the dwarf galaxies considered here. Stacking analyses have been used to detect X-ray emission in dwarf galaxies out to z $\approx$ 1.5 \citep{2016ApJ...817...20M, 2016ApJ...823..112P}. Additionally \cite{2016arXiv160301622P} used X-ray observations to search for AGN in dwarf galaxies at z $<1$, finding an AGN fraction of $\sim1\%$. However, we are concerned with following up individual systems in order to obtain a detailed look at the radiative properties of this relatively unexplored population.  

Determining the radiation properties of actively accreting BHs at the cores of dwarf galaxies is important for several reasons. Firstly, the BHs at the centers of dwarf galaxies may provide clues about galaxy nuclei in the early universe, since they are expected to be similar (to first approximation; see, e.g., \citealt{2011ApJ...742...13B, 2016arXiv160509394H}). 
With current instrumentation, it is not possible to detect $10^{5}M_{\odot}$ BHs in the earliest galaxies. A BH of this size accreting at its Eddington limit has a bolometric luminosity of $\sim10^{43}~{\rm erg~s^{-1}}$. Assuming it releases $\sim10\%$ of its energy in hard X-rays, the flux reaching us would be an order of magnitude below the detection limit of the 4 Ms Chandra Deep Field South (which has a 2-8 keV flux limit of $5.5\times10^{-17}~{\rm erg~s^{-1}~cm^{-2}}$; \citealt{2011ApJS..195...10X}). Furthermore, searches for AGN at high redshift (z $>6$) find fewer sources than expected based on relations at lower redshift \citep{2015MNRAS.448.3167W}, possibly due to the lower normalization for low-mass galaxies in the BH mass-galaxy stellar mass relation \citep{2015ApJ...813...82R, 2016ApJ...820L...6V}. As an alternative, present-day dwarf galaxies can serve as useful local analogs \citep{2011ApJ...731...55J, Reines:2011fr, 2014ApJ...787L..30R}. Present day dwarf galaxies have likely not undergone any major mergers, and are thus relatively undisturbed and ``pristine" compared to more massive systems. 

Moreover, studying AGN in dwarf galaxies is useful for understanding the interplay between AGN activity and star formation on all galaxy scales. AGN feedback is expected to have an effect on galaxy scale star formation, particularly in more massive systems (e.g., \citealt{2015ARAA..53..115K}). Feedback from massive stars and/or supernovae is expected to be particularly relevant for dwarf galaxies, but it is unclear what (if any) influence AGN can have on star formation in these smaller systems \citep{2005ApJ...618..569M, 2010MNRAS.401L..19H, 2014MNRAS.445..581H, 2016MNRAS.458..816H}. Studying radiation from AGN in dwarf galaxies is also useful for understanding whether BH accretion had any influence on star formation in the earliest galaxies (see e.g., \citealt{2012MNRAS.423.1325A}), as well as for investigating the contribution of low luminosity AGN to reionization \citep{2009ApJ...698..766M, 2015ApJ...813L...8M}. 

High resolution X-ray and UV follow-up of these systems is essential for understanding the accretion properties of AGN in dwarf galaxies. If detected, sufficiently bright, point-like nuclear UV/X-ray emission provides strong confirmation of the presence of an AGN (e.g., \citealt{1994ApJS...95....1E}). Additionally, X-ray studies can be used to determine the distribution of Eddington ratios for AGN in dwarf galaxies. Furthermore, the relative strength of the UV and X-ray emission is important for learning about the structure and properties of the accretion disk and corona (\citealt{1979ApJ...234L...9T,2016ApJ...819..154L}). Finally, studies of the broad-band spectra of these objects is necessary for determining the bolometric correction for this class of AGN. 

Reines et al. (2013) identified 151 dwarf galaxies with narrow and/or broad emission line signatures indicating the presence of an AGN. With the above goals in mind, we analyze \textit{Chandra X-ray Observatory} observations of a sub-sample of these objects, focusing on the most promising broad-line AGN candidates. The paper is organized as follows. In Section 2, we discuss our sample, X-ray and UV observations, and data reduction and analysis. In Section 3, we report on properties of the X-ray and UV emission, including the ratio of X-ray to UV emission. In Section 4, we discuss the origin of the X-ray emission, and compare the properties of our galaxies to more massive quasars.  

\section{Observations and Analysis}
\begin{figure}
\includegraphics[scale=0.6]{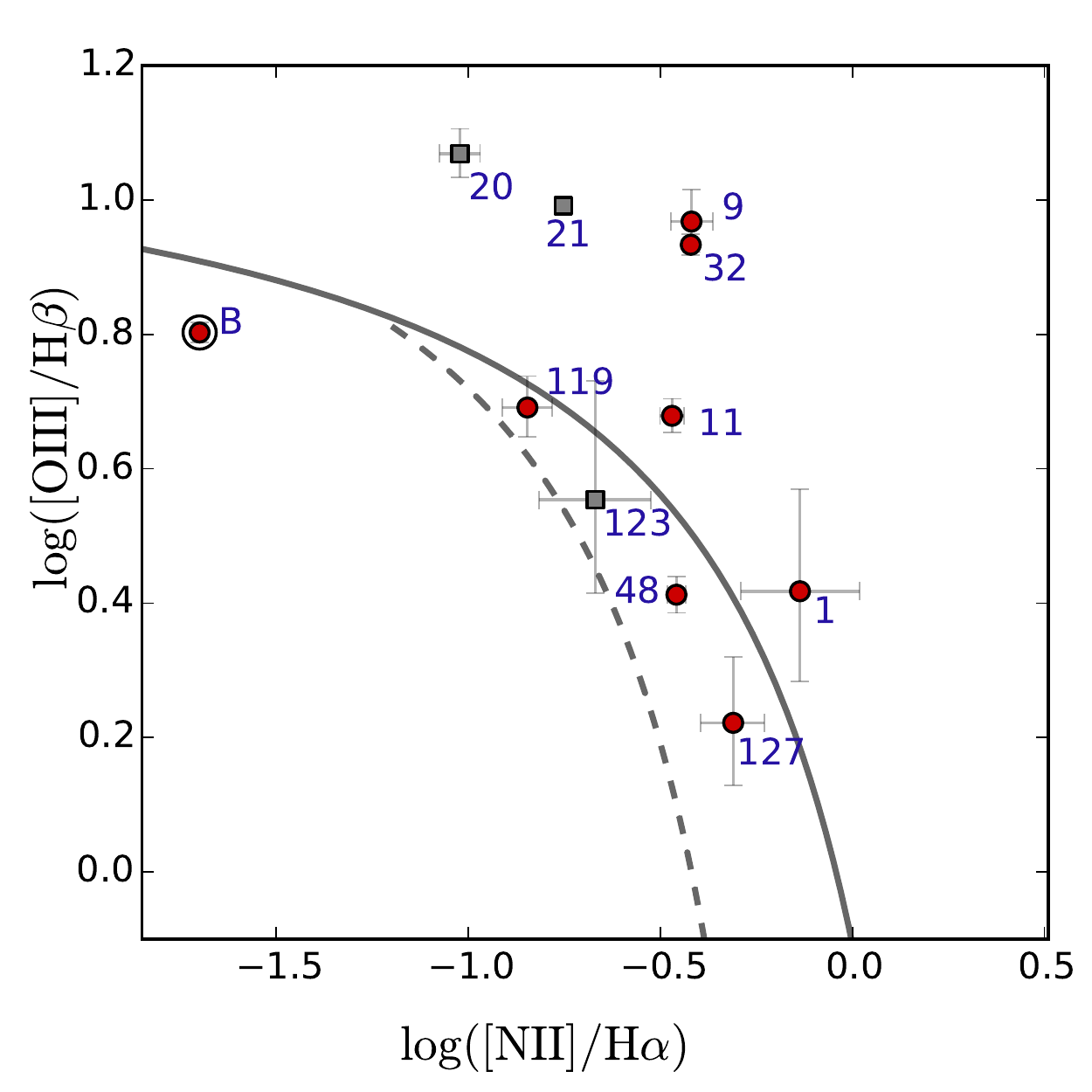}
\caption{Location of our targets on the BPT diagram. Red circles represent objects with new \textit{Chandra} observations, while gray squares represent objects with archival X-ray observations. The sole BPT star forming object is our sample is RGG B; in this plot and others in this paper, this object is surrounded by an open circle for clarity. \label{bpt} }
\end{figure}

Reines et al. (2013) identified 10 dwarf galaxies with both narrow \textit{and} broad emission line AGN signatures. Of these, six fall in the ``AGN" region of the BPT diagram, and four fall in the ``composite" region. Composite objects are expected to have contributions to narrow line emission from both an AGN and recent star formation. The 10 galaxies include the well-studied NGC 4395 (RGG\footnote{RGG refers to the ID assigned by Reines, Greene \& Geha (2013).}  21), as well as two objects (RGG 20 and 123) identified in the \cite{2007ApJ...670...92G} catalog and followed up in \cite{2012ApJ...761...73D} (see Section 2.3).  Figure~\ref{bpt} shows the location of our sample on the BPT diagram. 

We have obtained new \textit{Chandra X-ray Observatory} and \textit{Hubble Space Telescope} (HST) Wide-Field Camera 3 (WFC3) ultraviolet observations (Proposal ID: 16700103, PI: Reines) of the remaining seven. Additionally, we observed RGG B, a dwarf galaxy identified by Reines et al. (2013) to have broad H$\alpha$ emission but narrow-line ratios consistent with photoionization from HII regions. However, this object's narrow line ratios place it on the upper left of the star forming region of the BPT diagram, a regime where it is thought low-metallicity AGN can reside \citep{2006MNRAS.371.1559G}. In all, our full sample is comprised of 11 objects, three of which have \textit{Chandra} observations analyzed in the literature, and eight of which have new \textit{Chandra}/HST observations analyzed in this work. We stress that with our new observations, we complete high-resolution X-ray follow-up for all secure broad-line AGN identified in Reines et al. (2013).

\begin{figure*}
\centering
\includegraphics[scale=0.6]{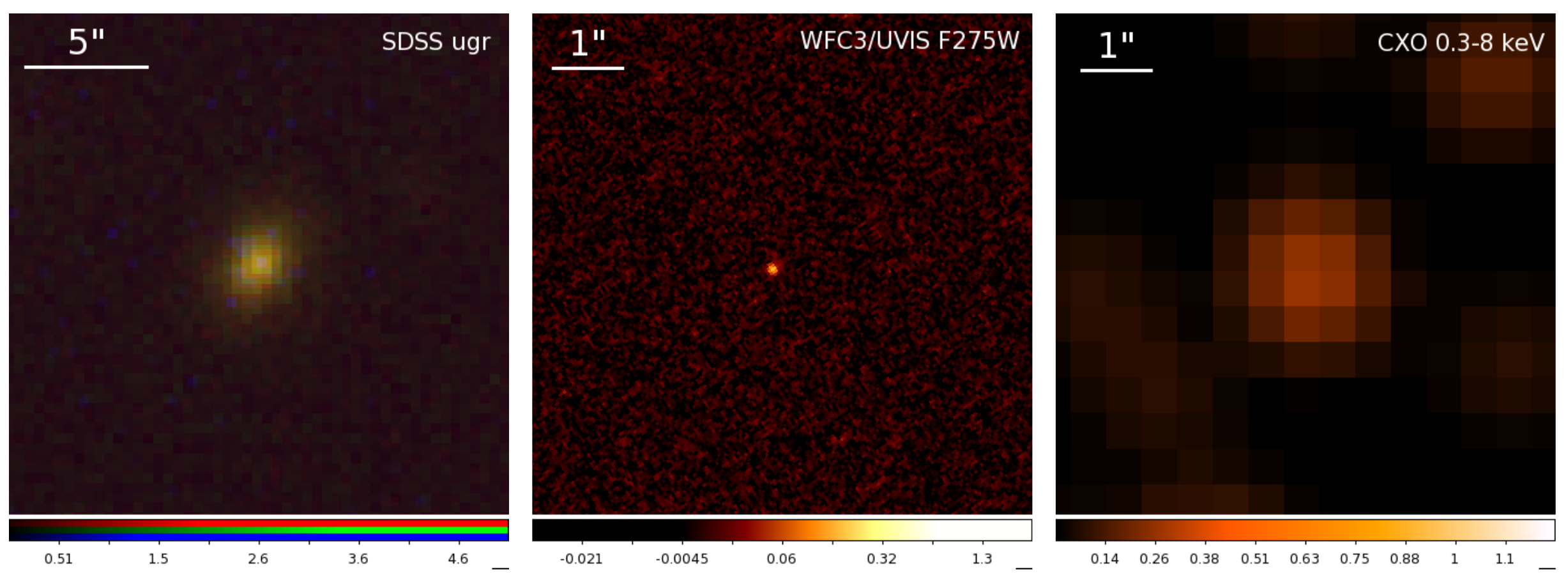}
\caption{Optical, UV, and X-ray imaging of RGG 1. The leftmost panel is a three-color SDSS image, using $u$, $g$, and $r$ bands as blue, green, and red, respectively. The center panel shows the HST WFC3/UVIS image taken with the F275W filter. The rightmost panel shows the \textit{Chandra X-ray Observatory} image, smoothed with a Gaussian kernel with radius of 3 pixels. At the median sample redshift of $z=0.04$, 1$''$ corresponds to $\sim850~{\rm pc}$. \label{rgg1}}
\end{figure*}

\begin{figure*}
\centering
\includegraphics[scale=0.6]{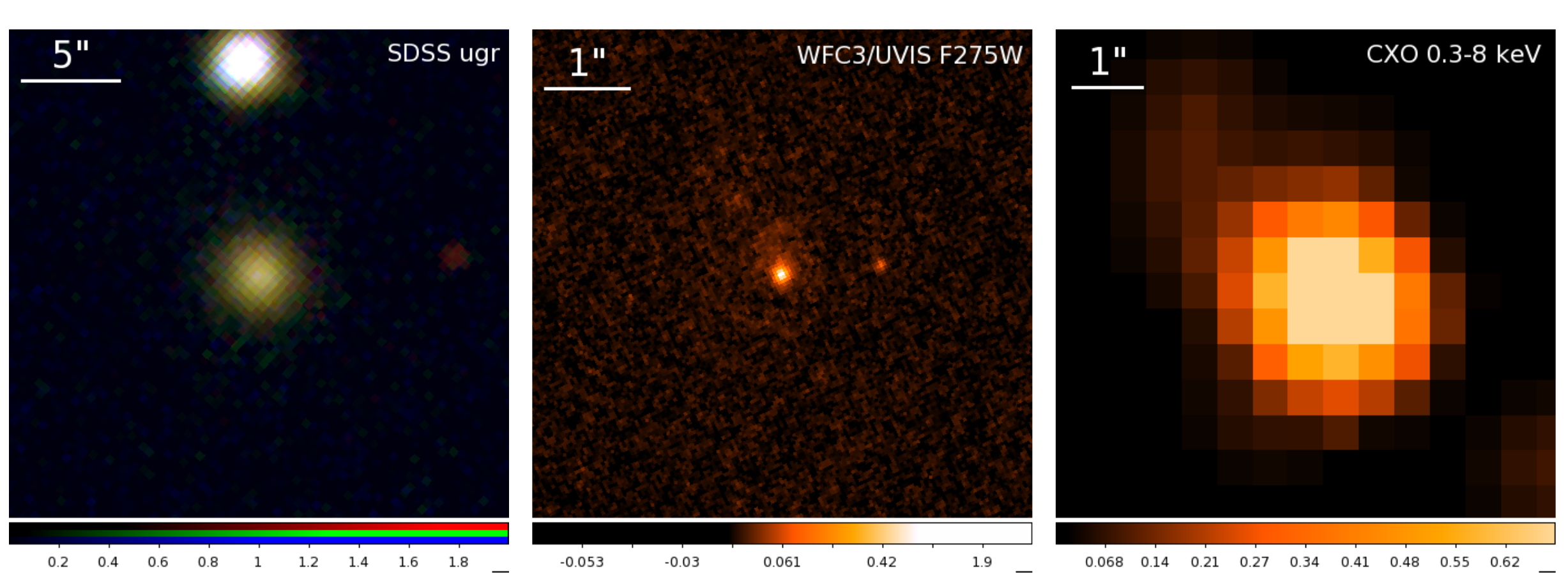}
\caption{Optical, UV, and X-ray imaging of RGG 9. See Figure~\ref{rgg1} for a more detailed description. \label{rgg9}}
\end{figure*}

\begin{figure*}
\centering
\includegraphics[scale=0.6]{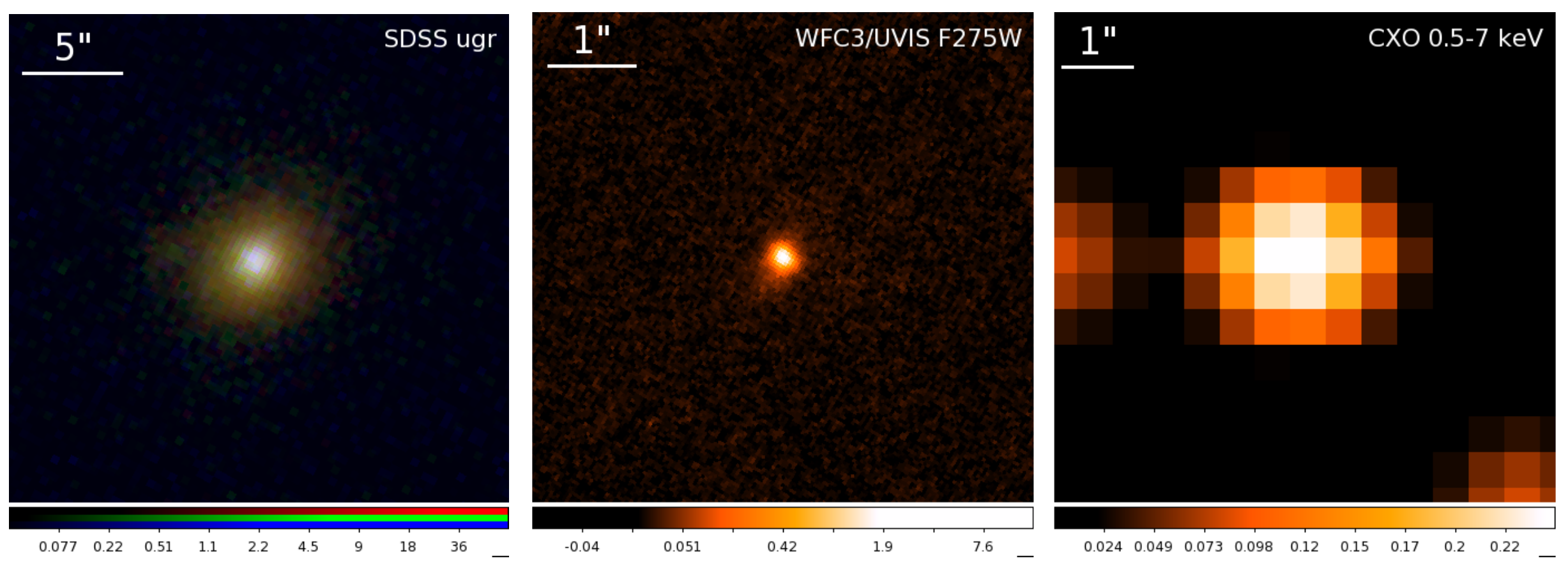}
\caption{Optical, UV, and X-ray imaging of RGG 11. See Figure~\ref{rgg1} for a more detailed description. \label{rgg11}}
\end{figure*}

\begin{figure*}
\centering
\includegraphics[scale=0.6]{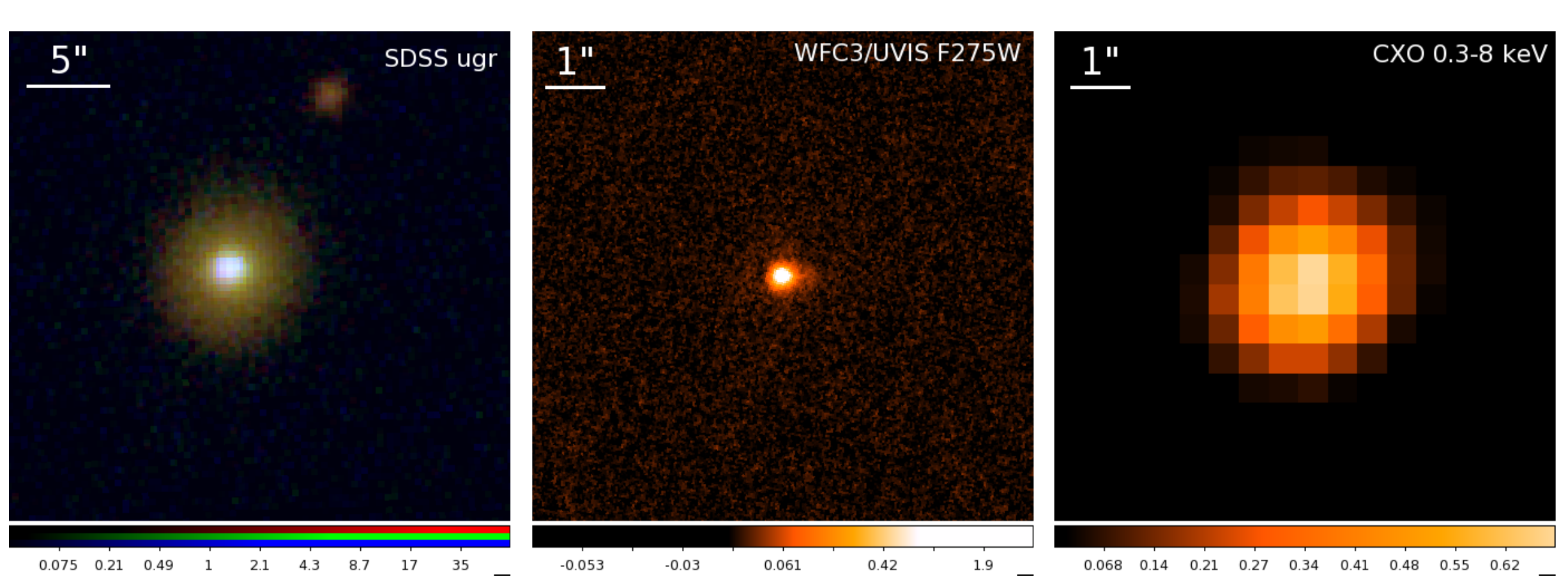}
\caption{Optical, UV, and X-ray imaging of RGG 32. See Figure~\ref{rgg1} for a more detailed description. \label{rgg32}}
\end{figure*}

\begin{figure*}
\centering
\includegraphics[scale=0.6]{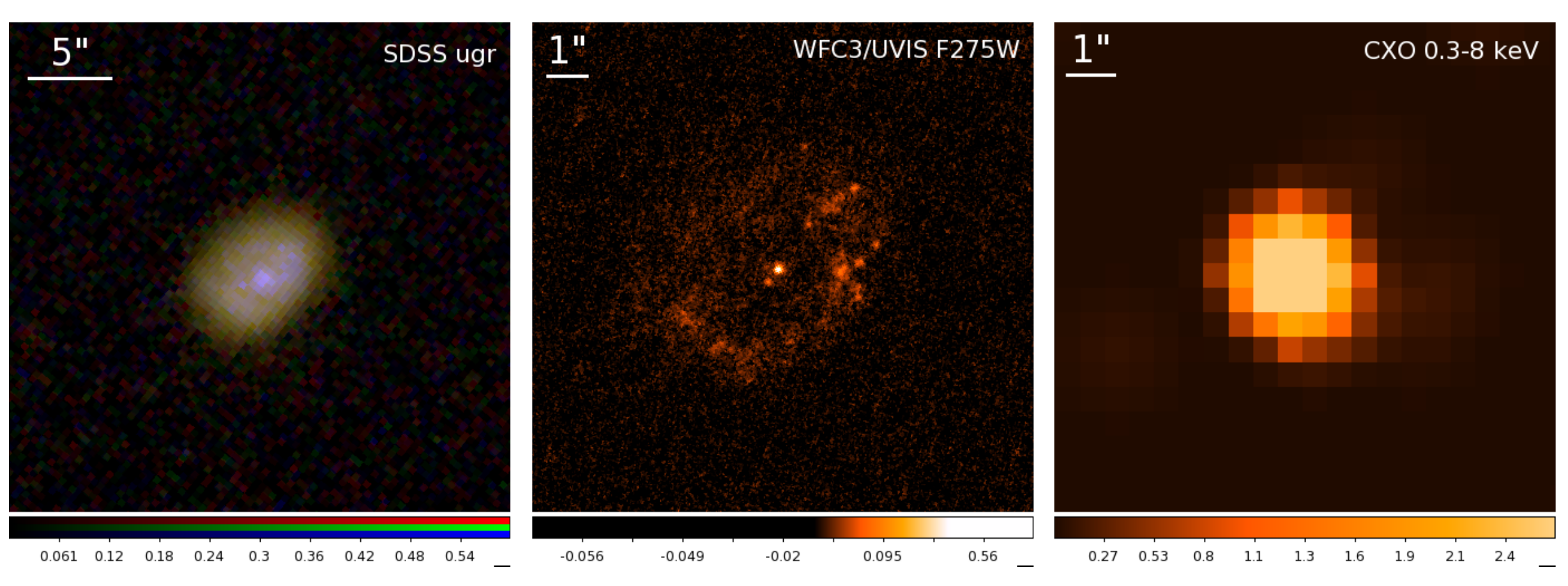}
\caption{Optical, UV, and X-ray imaging of RGG 48. See Figure~\ref{rgg1} for a more detailed description. \label{rgg48}}
\end{figure*}

\begin{figure*}
\centering
\includegraphics[scale=0.6]{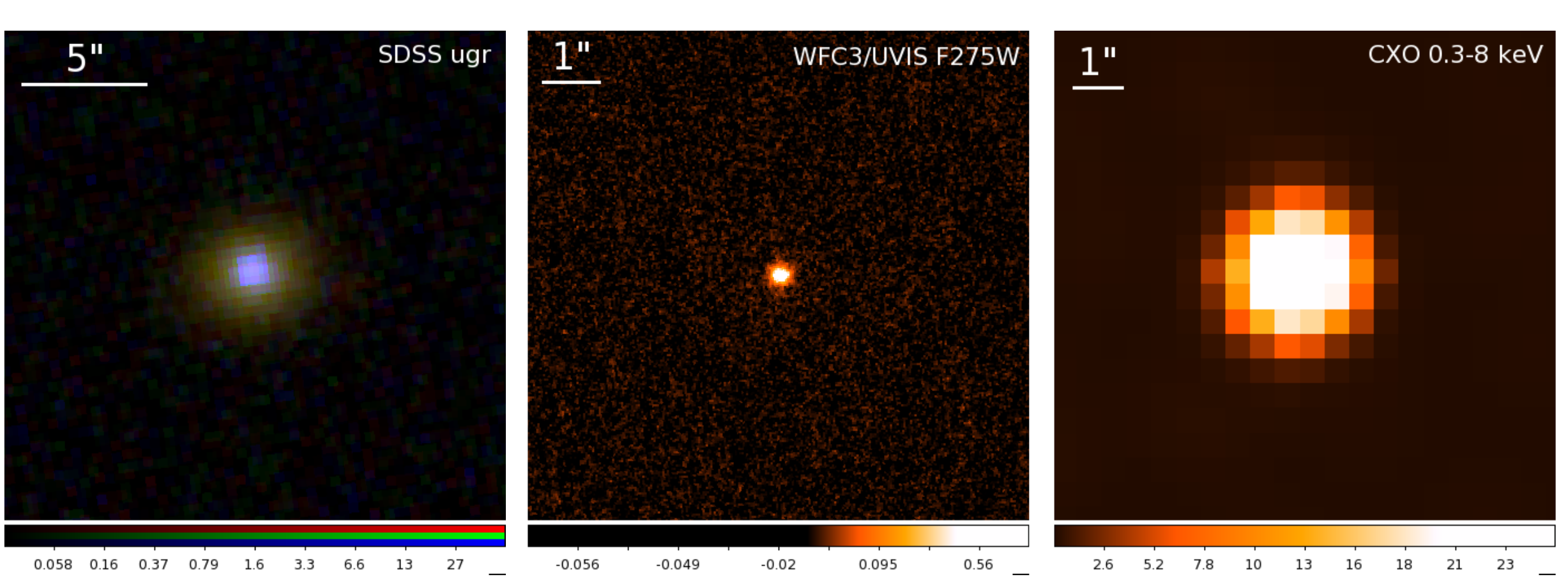}
\caption{Optical, UV, and X-ray imaging of RGG 119. See Figure~\ref{rgg1} for a more detailed description. \label{rgg119}}
\end{figure*}

\begin{figure*}
\centering
\includegraphics[scale=0.6]{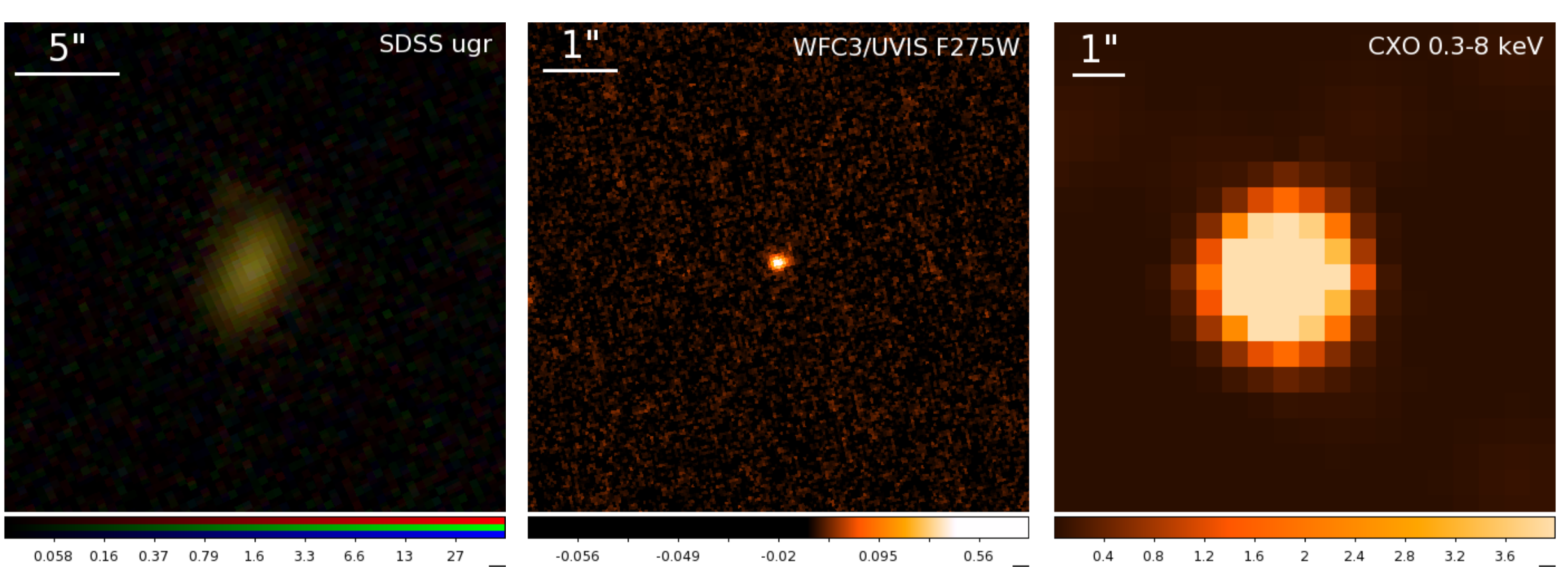}
\caption{Optical, UV, and X-ray imaging of RGG 127. See Figure~\ref{rgg1} for a more detailed description. \label{rgg127}}
\end{figure*}

\begin{figure*}
\centering
\includegraphics[scale=0.6]{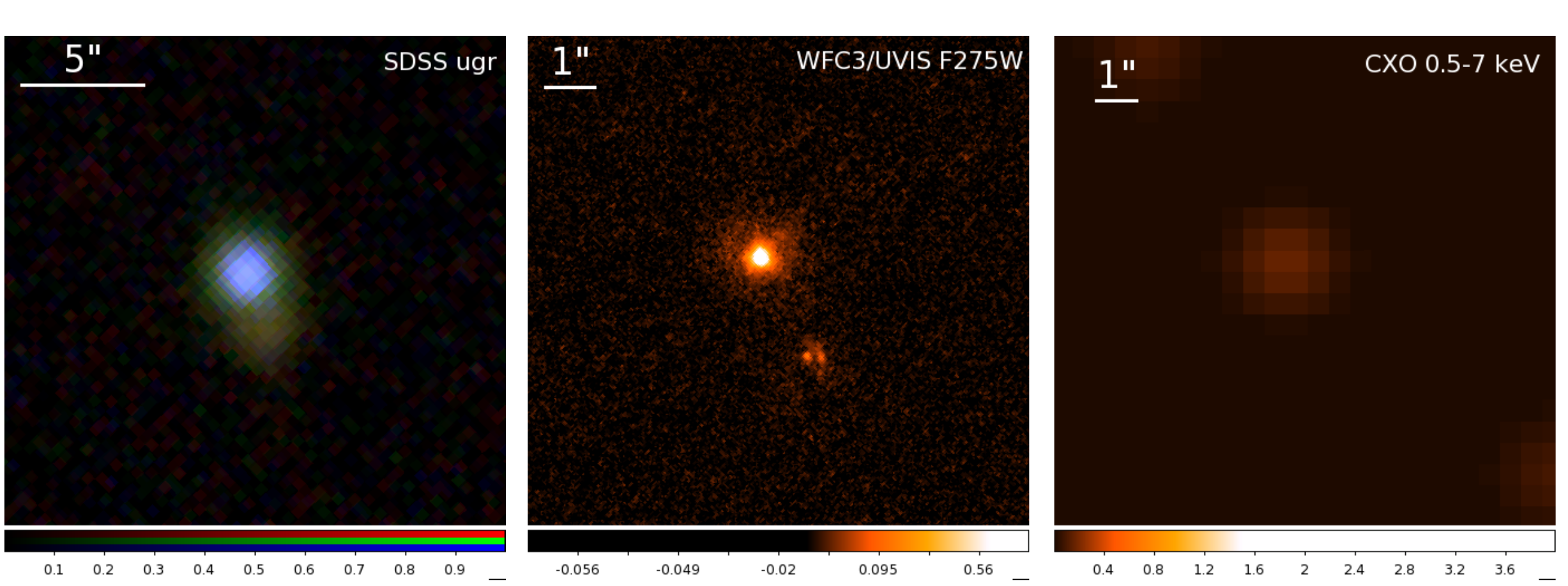}
\caption{Optical, UV, and X-ray imaging of RGG B. See Figure~\ref{rgg1} for a more detailed description. Note that RGG B is the sole object with narrow line ratios placing it in the star formation region of the BPT diagram.}
\label{rggb}
\end{figure*}

All targets are nearby ($z<0.055$), and stellar masses range from $\sim1\times10^{8}-3\times10^{9}~M_{\odot}$ (see Tables 1 and 3 in \citealt{Reines:2013fj}). Figures~\ref{rgg1} through \ref{rggb} show, for each of our eight new targets, the three-color SDSS image, the HST UV image, and the CXO image.

\subsection{Chandra X-ray Observatory}

\textit{Chandra X-ray Observatory} observations of our targets were taken with the Advanced CCD Imaging Spectrometer S-array (ACIS-S) between December 2014 and April 2016 (PI: Reines, Cycle 16). See Table~\ref{obs} for a summary of observations. The data were reprocessed and analyzed with the \cxo Interactive Analysis of Observations software package (CIAO; version 4.8). After reprocessing and restricting our data to the S3 chip, we generated a preliminary source list using CIAO WAVDETECT, and corrected the \cxo astrometry by cross-matching our observations with sources in the Sloan Digital Sky Survey catalog (excluding our own). In 3 observations (RGG 11, 32, B), there were not a sufficient number of cross-matched sources to apply an astrometry correction. For sources where there were three or more cross-matched sources, astrometric corrections ranged from $\sim0.2-0.6''$.  We checked all observations for background flares; none were found. We then applied an energy filter (0.3-8 or 0.5-7 keV) and re-ran WAVDETECT using a threshold significance of $10^{-6}$ (equivalent to 1 false detection over the S3 chip). 

We used SRCFLUX to calculate count rates and errors in three energy bands: soft ($0.5-2$ keV), hard ($2-10$ keV), and broad ($0.5-7$ keV). For targets with a WAVDETECT source coincident with the optical nucleus (RGG 9, 32, 48, 119, 127), we extracted counts in a circular source region of radius 2$''$ centered on the WAVDETECT coordinates. In the absence of a WAVDETECT source (RGG 1, 11, B), we extracted counts in a region of the same size centered on the nominal position of the nucleus as given by SDSS coordinates. We used a background annulus with inner and outer radii of 20 and 30$''$, respectively. All sources were detected in the soft and broad bands. 

Counts were converted to unabsorbed fluxes using the Portable, Interactive Multi-Mission Simulator (PIMMS)\footnote{http://cxc.harvard.edu/toolkit/pimms.jsp}. We assumed the underlying spectrum to be a power law with $\Gamma=2.0$, except in the cases of RGG 119 and RGG 127, for which we were able to measure $\Gamma$ directly from the extracted spectra (see below). Galactic HI column density measurements were obtained with the HEASARC nH tool\footnote{http://heasarc.nasa.gov/cgi-bin/Tools/w3nh/w3nh.pl}, which uses HI column densities measured by \cite{2005AA...440..775K}. Table~\ref{counts} presents the counts, unabsorbed fluxes, and luminosities for each target in each wave band. 

\begin{figure*}
\includegraphics[width=6.5in]{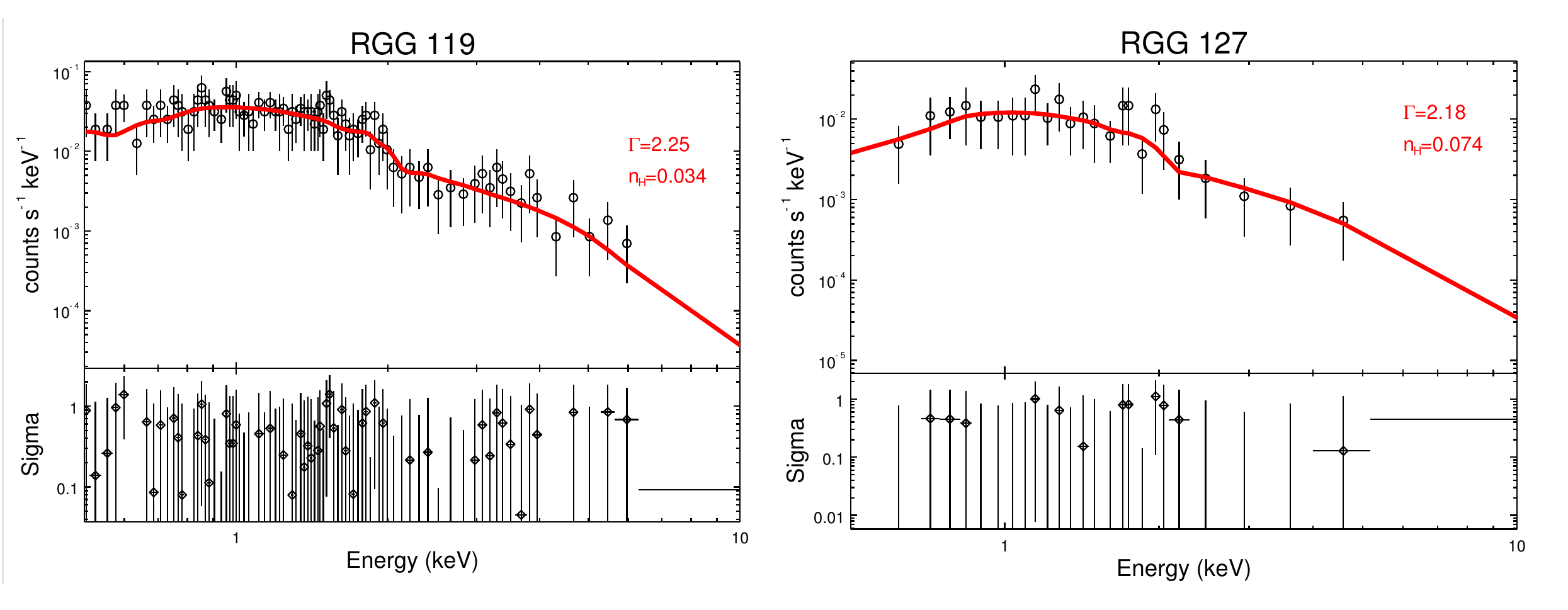}
\caption{\textit{Left}: X-ray spectrum of RGG 119. We fit the spectrum with an absorbed power law model, finding a best fit $\Gamma=2.25$. \textit{Right}: X-ray spectrum of RGG 127. We fit the spectrum with an absorbed power law model, finding a best fit $\Gamma=2.18$ and $n_{H}=7.4\times10^{21}~\rm{cm^{-2}}$ (approximately twice the galactic $n_{H}$ value).}
\label{spec}
\end{figure*}

Two sources (RGG 119 and 127) had enough counts (n $\gtrsim100$) to warrant spectral analysis (Figure~\ref{spec}). We use the Sherpa fitting package in CIAO to model the extracted spectra using an absorbed power law model. Since it is possible to have both Galactic absorption and absorption intrinsic to the galaxy itself, we fit the spectra two ways: with $n_{H}$ frozen to the Galactic value, and with $n_{H}$ constrained to be at least the Galactic value. We find that the spectrum of RGG 119 has a best fit power law index of $\Gamma=2.25\pm0.25$. The models did not prefer an $n_{H}$ greater than the galactic value, suggesting little-to-no intrinsic absorption. RGG 127, however, has a best fit $n_{H}$ approximately twice the galactic value, indicating potential intrinsic absorption. For RGG 127, we find a corresponding best fit $\Gamma$ of $2.18^{+0.55}_{-0.35}$. If we freeze $n_{H}$ to the galactic value, we obtain $\Gamma=2.03$.

\subsection{Hubble Space Telescope}

\textit{HST} WFC3 images were taken with the UVIS channel and F275W filter. Individual exposures were processed and combined using the standard HST AstroDrizzle reduction pipeline. Total exposure times ranged from $747~\rm{s}$ to $900~\rm{s}$. We detect UV emission from all eight targets. 

We use the PHOT task within the STSDAS package for PyRAF to measure UV fluxes. The UV emission associated with the AGN emerges from the unresolved accretion disk. There may also be spatially resolved UV emission on scales larger than the WFC3 PSF. In order to isolate emission from the AGN, we measure the flux within an aperture with a radius of 3 pixels centered on the brightest pixel, accounting for the HST WFC3/UVIS F275W encircled energy fraction. We try four different background apertures ranging from an annulus immediately surrounding the source (effectively subtracting off any nearby star formation) to a large annulus containing only the sky background. The choice of background aperture does not strongly affect our results, but since we do expect some UV emission from star formation in addition to any from the AGN, we use the innermost annulus background for our measurements of the UV-to-X-ray flux ratio. Table~\ref{uv} presents measured count rates for the background and AGN.

\subsection{Archival data}

{\textit{Chandra} imaging of RGG 21 (NGC 4395) was analyzed in \cite{2005AJ....129.2108M}, and RGG 20 and RGG 123 were analyzed in \cite{2012ApJ...761...73D}. We briefly summarize their findings here. 

\textit{RGG 20}. RGG 20 (SDSS J122342.82+581446.4) was found to have a 0.5-8 keV luminosity of $L=6.7\times10^{41}~\rm{erg~s^{-1}}$. There were a sufficient number of counts to extract a spectrum; the best-fit photon index was $\Gamma = 1.54\pm0.10$. The ratio of X-ray to UV emission ($\alpha_{\rm{OX}}$; see Section 3.2) is $\alpha_{\rm{OX}}$= -1.30.

\textit{RGG 21}. RGG 21 (NGC 4395; SDSS J122548.86+333248.7) was found to have a 2-10 keV luminosity of $L=8.0\times10^{39}~\rm{erg~s^{-1}}$. NGC 4395 displays dramatic spectral variability, with measured power law indices ranging from $\Gamma = 1.72$ \citep{2000MNRAS.318..879I} to $\Gamma = 0.61$ \citep{2005AJ....129.2108M} on timescales of several years.  

\textit{RGG 123}. RGG 123 (SDSS J153425.58+040806.6) has a 0.5-8 keV luminosity of  $L=8.5\times10^{41}~\rm{erg~s^{-1}}$. There were not sufficient counts to extract an X-ray spectrum. The photon index as estimated from the hardness ratio is $\Gamma_{\rm HR} = 2.57\pm0.27$\footnote{\cite{2012ApJ...761...73D} found that the photon index estimated from the hardness ratio is highly correlated with the photon index of the fitted spectrum for $\Gamma>2$.}, and $\alpha_{\rm{OX}}$ was found to be -1.22. 

\cite{2012ApJ...761...73D} reports count rates for RGG 20 and 123 (as well as fluxes and luminosities) in the 0.5-2 keV and 2-8 keV bands. For uniformity, we convert their 2-8 keV count rates to 2-10 keV fluxes using PIMMS. For RGG 20, we use the photon index of $\Gamma = 1.54$ obtained from the fit to the spectrum. For RGG 123, we follow the same procedure as for our new data and use a photon index of $\Gamma = 2.0$. }

\section{Results}

\subsection{Nuclear X-ray Emission}

All eleven targets are detected, i.e. we find a 100\% X-ray detection rate for nearby ($z<0.055$) dwarf galaxies with broad and narrow-line AGN signatures (i.e., those falling in the AGN \textit{or} composite region of the BPT diagram). For our 8 new targets, we measure count rates in the $0.5-7$ keV band ranging from $1.5\times10^{-4}~\rm{cts/s}$ to $5.2\times10^{-2}~\rm{cts/s}$. Correspondingly, we estimate $0.5-7$ keV luminosities ranging from $\sim5\times10^{39}$ to $\sim1\times10^{42}~\rm{erg~s^{-1}}$. Distances are estimated using redshifts reported in the NASA-Sloan Atlas and Ned Wright's cosmology calculator \citep{2006PASP..118.1711W}. 

If we assume that the measured X-ray luminosities are due to accreting BHs (see Section 4.1 for a discussion of the origin of X-ray emission for these sources), we can estimate Eddington fractions, i.e. at what percentage of their Eddington luminosities these BHs are accreting. Eddington luminosities are determined using the BH masses from Reines et al. (2013), with the exceptions of RGG 9 and RGG 119, which have updated BH masses given in \cite{2016arXiv160505731B}. 
BH masses for this sample are estimated using the luminosity and FWHM of the broad H$\alpha$ emission line \citealt{2000ApJ...533..631K, 2004ApJ...613..682P, 2005ApJ...630..122G,2009ApJ...705..199B, 2013ApJ...767..149B} and range from $7\times10^{4} - 1\times10^{6}~M_{\odot}$ (\citealt{Reines:2013fj}, \citealt{2016arXiv160505731B}). We caution that bolometric corrections for AGN in this mass regime are largely unconstrained. However, in the absence of a full analysis of the spectral energy distribution of AGN in dwarf galaxies, we use a hard X-ray bolometric correction of $L_{\rm{Bol}}/L_{\rm{2-10 keV}} = 10$ \citep{2004MNRAS.351..169M}. For $L_{\rm bol} = 10~L_{\rm 2-10 keV}$, we find Eddington fractions ranging from $\sim0.1-50$\%, including the archival observations (see Table~\ref{accretion} for Eddington ratios). The median Eddington fraction for this sample of 10 dwarf AGN is $4.3\%$ (mean $\approx10\%$); this range is comparable to the range of Eddington ratios for broad-line quasars at higher redshifts (see, e.g., \citealt{Kelly:2010qy, 2012ApJ...746..169S}). Note that we have excluded RGG B from this analysis, as we do not consider it a secure AGN; see Section 4.1.

\subsection{UV-to-X-ray flux ratios}

We measure $\alpha_{\rm{OX}}$ in order to quantify the relative power output in the UV and in X-rays \citep{1979ApJ...234L...9T}. The quantity $\alpha_{\rm{OX}}$ is defined as $\alpha_{\rm{OX}} = -0.383 \log(l_{2500}/l_{2\rm{keV}})$, where $l_{2500}$ is the luminosity density at $2500\rm{\AA}$ and $l_{2\rm{keV}}$ is the unabsorbed luminosity density at 2 keV. We measure the $2500\rm{\AA}$ UV flux density directly from the HST F275W imaging (the peak of the filter output is close to $2600\rm{\AA}$, i.e. $\sim2500\rm{\AA}$ for a galaxy at z=0.04). We use PIMMS and the appropriate power-law index ($\Gamma=2.0$, except when measured directly from spectral fitting) to determine the $2\rm{keV}$ luminosity. 

\begin{figure}
\includegraphics[scale=0.6]{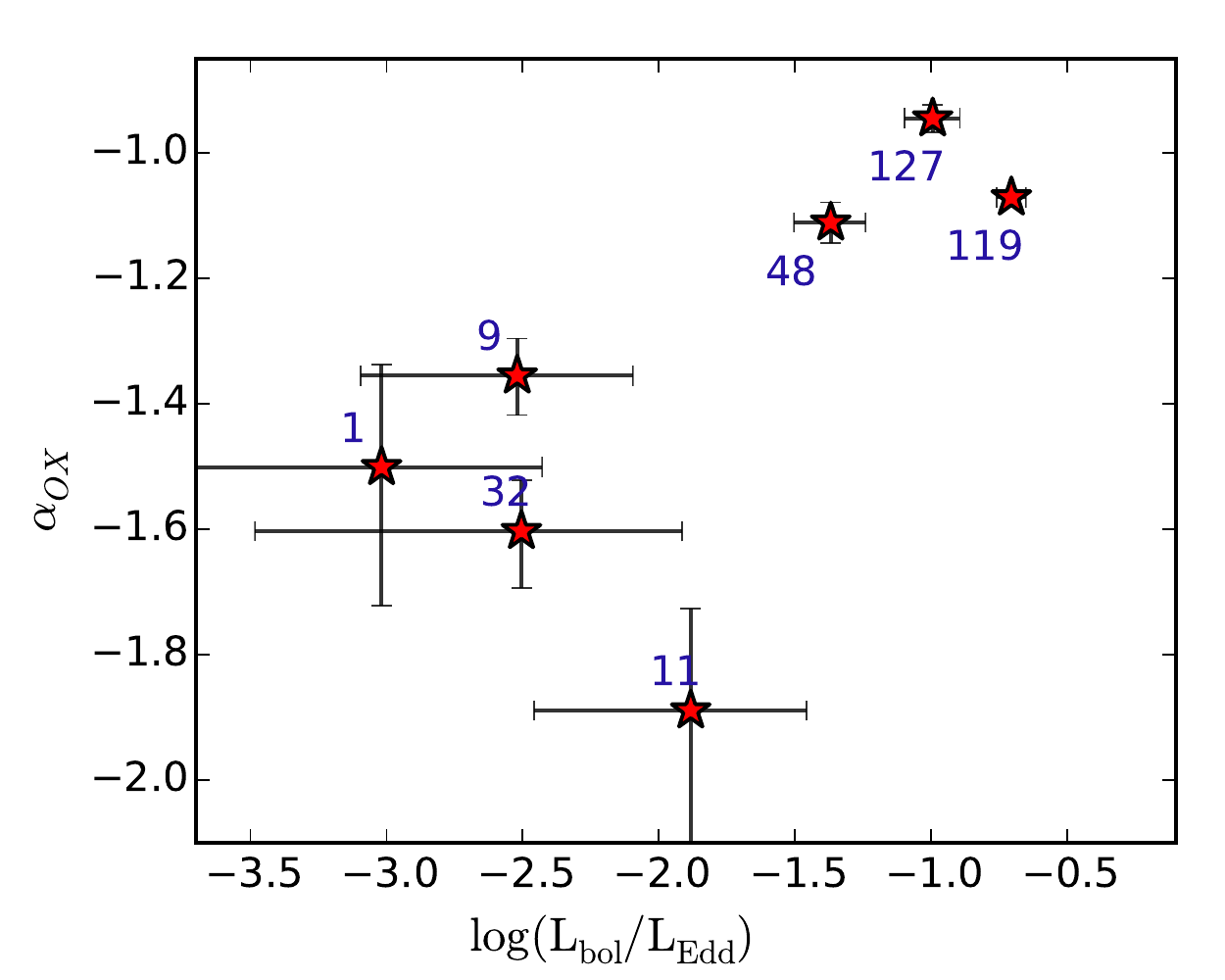}
\caption{Ratio of UV-to-X-ray emission, $\alpha_{\rm{OX}}$, versus Eddington ratio. $\rm{L_{bol}/L_{Edd}}$ is calculated using a bolometric correction of 10 \citep{2004MNRAS.351..169M}.\label{alpha_ox}}
\end{figure}

We find $\alpha_{\rm{OX}}$ values ranging from $-2.13$ to $-0.95$ (Figure~\ref{alpha_ox}). These values are consistent with those found by  \cite{2012ApJ...761...73D} and \cite{2016ApJ...825..139P} for low-mass AGN with BH masses of $\sim10^{6}M_{\odot}$.  Though we have a very limited sample size, broadly speaking, the highest $\rm{L_{bol}/L_{Edd}}$ objects have the least negative $\alpha_{\rm{OX}}$ values, i.e. are more X-ray luminous relative to their UV emission. 

\begin{figure}
\includegraphics[scale=0.6]{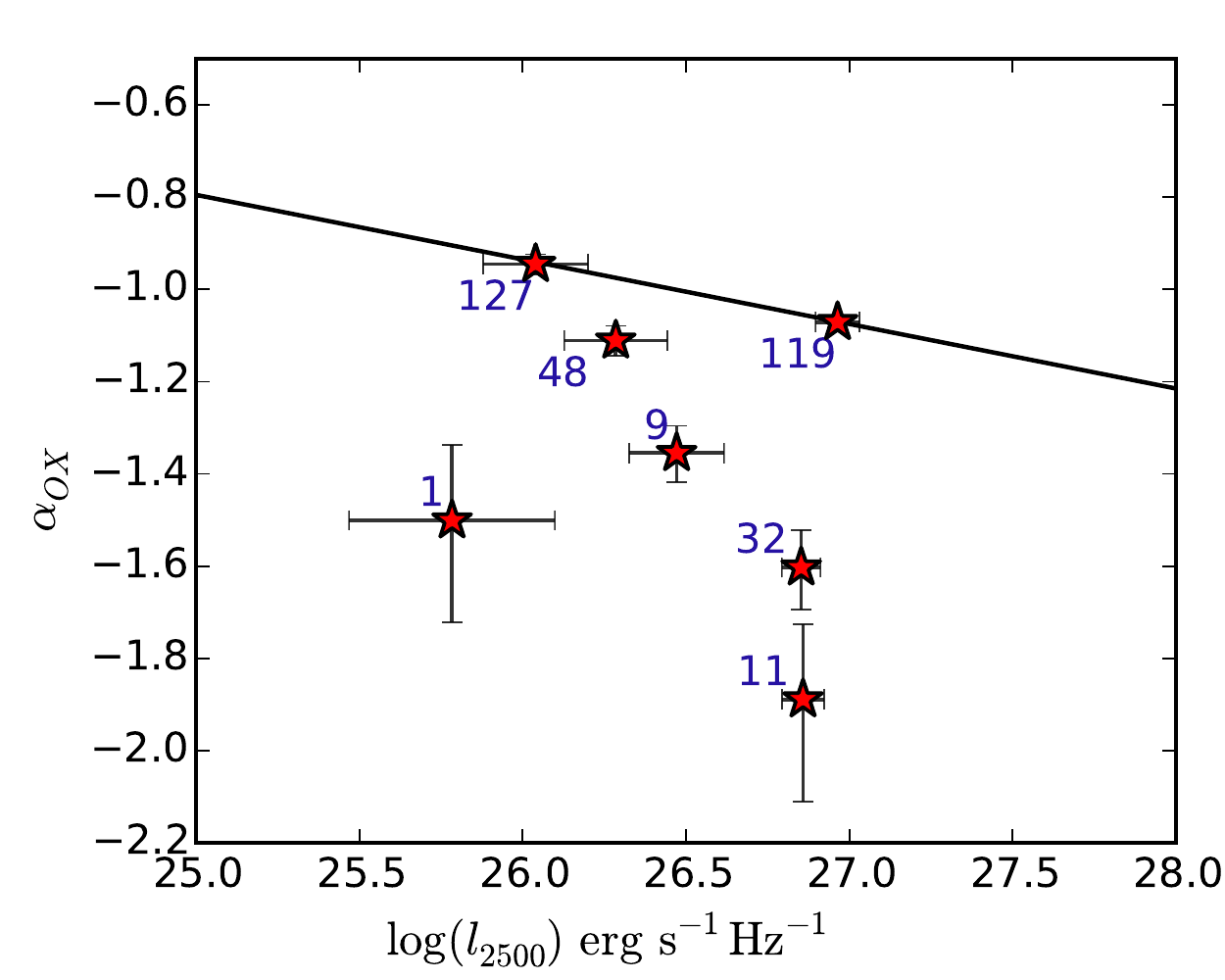}
\caption{Ratio of UV-to-Xray emission ($\alpha_{\rm{OX}}$) versus luminosity at 2500 $\rm{\AA}$. The solid black line represents the relationship between $l_{2500}$ and $\alpha_{\rm{OX}}$ found by \cite{2007ApJ...665.1004J} using a sample of 372 quasars from $z=1.5 - 4.5$. \label{l2500_alpha_ox}}
\end{figure}

In Figure~\ref{l2500_alpha_ox}, we compare our $\alpha_{\rm{OX}}$ values to expected values based on the relationship between $\alpha_{\rm{OX}}$ and $l_{2500}$ found by \cite{2007ApJ...665.1004J}.  We find that our targets tend to have $\alpha_{\rm{OX}}$ \textit{lower} than expected based on the $\alpha_{\rm{OX}}$-$l_{2500}$
relation (the $\alpha_{\rm OX}$ values range from $\sim0.1-1$ below the expected values). Despite the general trend, RGG 119 and RGG 127, i.e. the two objects with the highest $\rm{L_{bol}/L_{Edd}}$, have $\alpha_{\rm{OX}}$ consistent with their ``expected" values. The large dispersion in $\alpha_{\rm{OX}}$ values for BHs with $M_{\rm BH}\lesssim10^{6} M_{\odot}$, as well as their tendency to appear X-ray weak relative to the $\alpha_{\rm{OX}}-l_{2500}$ relation is also observed by  \cite{2012ApJ...761...73D} and \cite{2016ApJ...825..139P}.  This is further discussed in Section 4.2.  

We note that it is very difficult to distinguish emission from the AGN from emission from star formation. If a significant fraction of the UV emission is from star formation, then $\alpha_{\rm{OX}}$ values would appear \textit{lower} (i.e., more negative) than they are. 

\subsection{X-ray hardness ratio}

\begin{figure}
\includegraphics[scale=0.6]{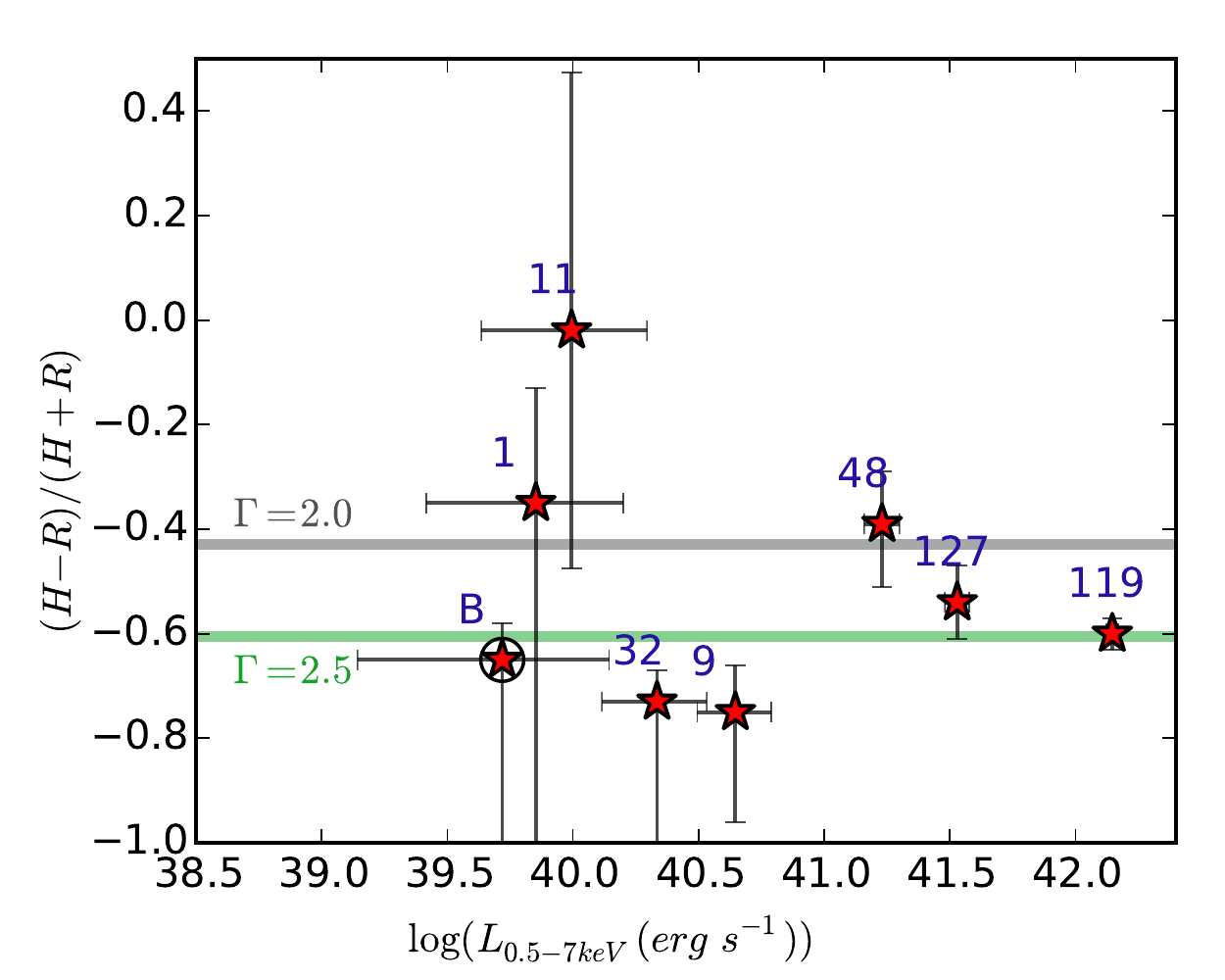}
\caption{Hardness ratio versus 0.5-7 keV luminosity. Hardness ratios were computed using the BEHR code; error bars represent 90\% confidence intervals. For reference, we show hardness ratios computed using the WEBPIMMS tool for unabsorbed power laws with $\Gamma=2.0$ and $\Gamma=2.5$. The lone BPT star forming object (RGG B) is also marked with an open circle. }
\label{hardness}
\end{figure}

Hardness ratios can yield information about the spectral shape for sources with too few counts to extract a spectrum. We compute hardness ratios for our targets using the Bayesian Estimation of Hardness Ratios code (BEHR; \citealt{2006ApJ...652..610P}). Hardness ratio is defined as $\rm{(H-S)/(H+S)}$, where H and S are the number of counts in the hard and soft bands, respectively. Here, we use 0.5-2 keV as our soft band and 2-7 keV as the hard band. BEHR treats counts as independent Poisson random variables, and can be used even if the source is undetected in one of the bands (as is the case for RGG B). We find hardness ratios ranging from $-0.75$ to $-0.02$, with a median hardness ratio of $-0.57$ (see Figure~\ref{hardness}). For reference, we compute expected hard and soft band counts for power law spectra with varying $\Gamma$; a hardness ratio of $\sim-0.4$ is characteristic of an unabsorbed power law spectrum with $\Gamma=2.0$, while an unabsorbed power law spectrum with $\Gamma=2.5$ produces a hardness ratio of $\sim-0.6$. 

The hardness ratios for RGG 119 and RGG 127 are consistent with the expected hardness ratios for $\Gamma\approx2.0-2.5$; these are in turn consistent with the $\Gamma$ which best fit the extracted X-ray spectra (2.25 and 2.18 for RGG 119 and RGG 127, respectively).

\section{Discussion}
\subsection{Origin of X-ray emission}

The Eddington luminosity\footnote{$L_{\rm{Edd}} = (\rm{M_{BH}/M_{\odot}})\times1.25\times10^{38}~\rm{erg}~\rm{s^{-1}}$} for a $10^{5}~M_{\odot}$ black hole is $\sim10^{43}$ \ergs. Therefore, the X-ray luminosities of Eddington-limited X-ray binaries can be comparable to the luminosities of sub-Eddington AGN in our BH mass range of interest. Here, we carefully consider the origin of the detected X-ray emission in our targets. 

The expected X-ray luminosity due to high mass X-ray binaries (HMXBs) scales with the star formation rate of the galaxy \citep{2003MNRAS.339..793G}, while the X-ray luminosity due to low-mass X-ray binaries (LMXBs) scales with stellar mass \citep{:ut}. As shown in \cite{:ut}, for a galaxy with $M_{\ast}\sim10^{9}M_{\odot}$, we expect LMXBs to contribute $L_{X}<10^{38}$\ergs; our X-ray sources are all more luminous by at least a factor of 10, making LMXBs an unlikely source of the nuclear X-ray emission. Additionally, while the fairly soft best-fit spectral indices for RGG 119 and 127 are typical of quasars, they are also typical of high mass X-ray binaries (HMXBs), particularly in the steep power law and high/soft states \citep{2006ARA&A..44...49R}.
Thus, we consider HMXBs our most likely alternative source of X-ray emission. We estimate the expected luminosity contributed from HMXBs below. 

\begin{figure*}
\centering
\includegraphics[scale=0.63]{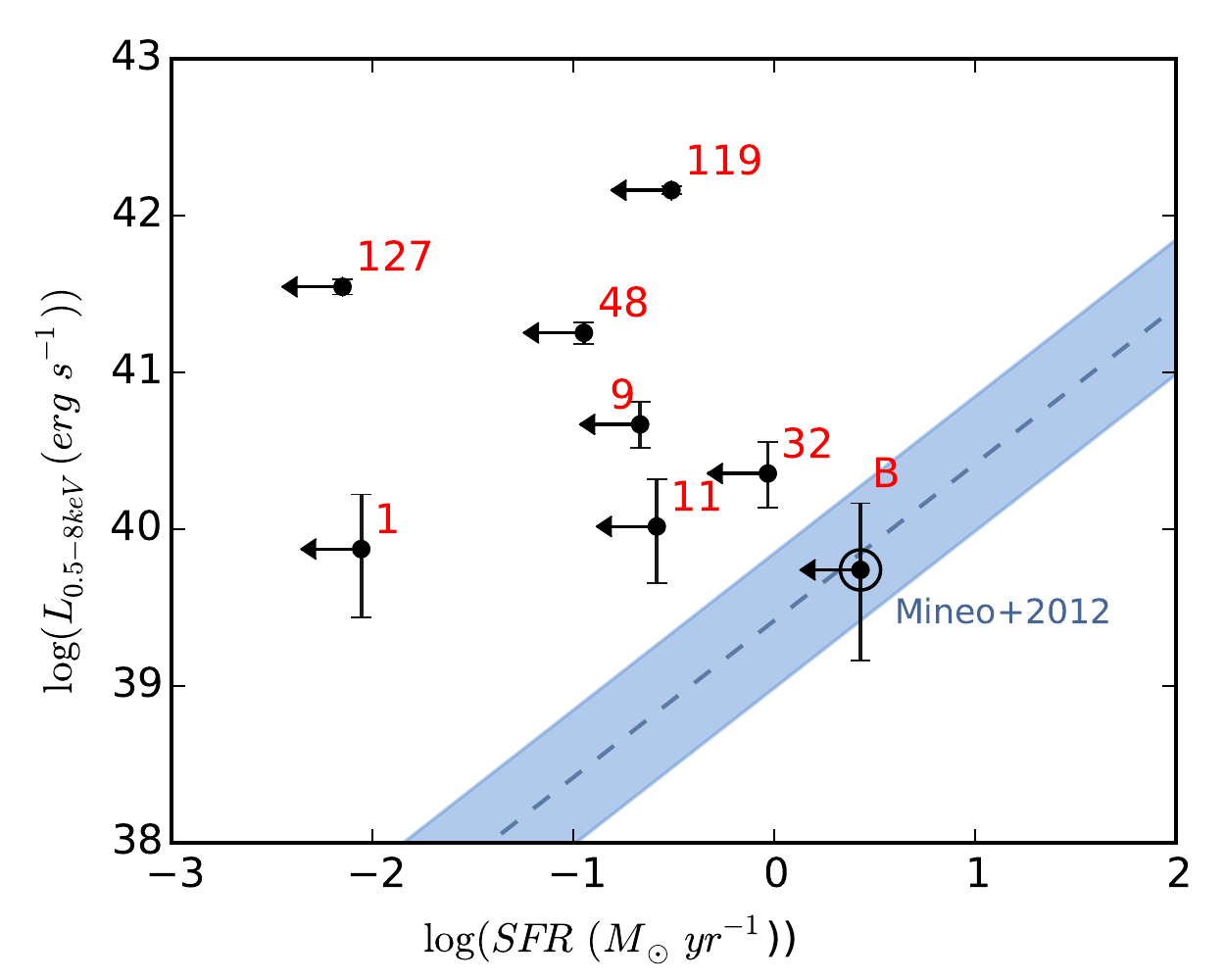}
\includegraphics[scale=0.63]{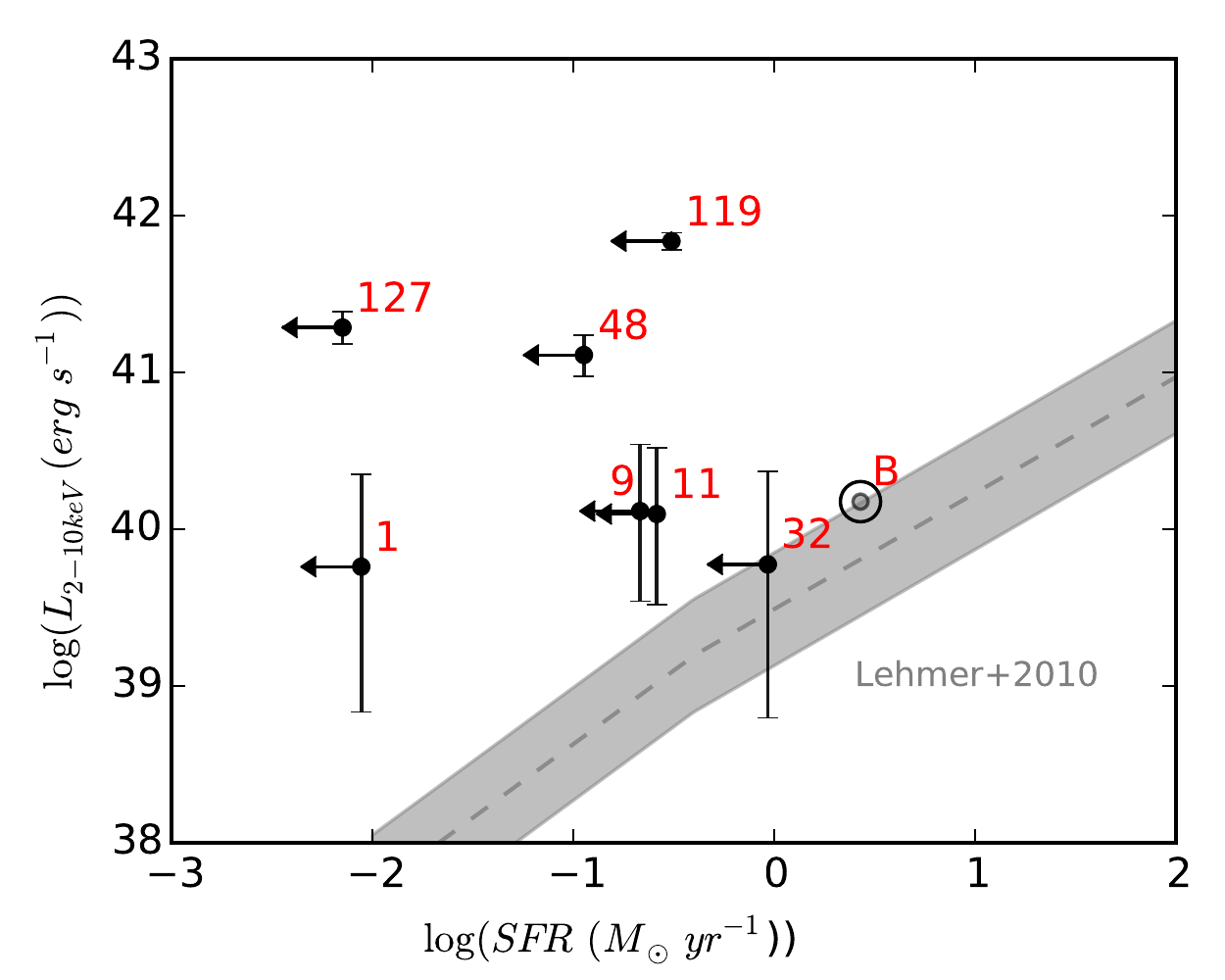}
\caption{X-ray luminosity versus SFR for our eight targets. SFRs are conservative upper-limits determined based on the luminosity of narrow H$\alpha$ (see text). The left panel uses 0.5-8 keV luminosities and the relation defined in Mineo et al. (2012). The right panel uses 2-10 keV luminosities and the relation defined by Lehmer et al. (2010). On the right panel, RGG B, the BPT star forming object, has only an upper limit on the 2-10 keV luminosity, and is shown as a faded point. In both panels, RGG B is also marked with an additional open circle.}
\label{lx_v_sfr}
\end{figure*}

Using the extinction-corrected luminosity of the narrow H$\alpha$ emission from \cite{Reines:2013fj} and the relation in \cite{2012ARAA..50..531K}, we compute upper limits on the SFR for each galaxy by assuming that all the narrow H$\alpha$ emission within the SDSS fiber is due to star formation. Note that for objects which fall in the composite or AGN region of the BPT diagram, we can be confident there is some contribution to the H$\alpha$ emission from the AGN. Also note that the 3$''$ SDSS fiber is larger than the 2$''$ \textit{Chandra} PSF, so we are calculating the expected contribution from HMXBs over a slightly larger area than necessary. We use relations from \cite{2010ApJ...724..559L} and \cite{2012MNRAS.419.2095M} to compute, respectively, the expected 2-10 keV and 0.5-8 keV luminosities from HMXBs. As shown in the left panel of Figure~\ref{lx_v_sfr}, the 0.5-8 keV luminosities of all targets with broad \textit{and} narrow optical emission line AGN signatures are higher (by $1.5-18\sigma$) than would be expected from star formation alone. RGG B, the sole target with narrow line ratios consistent with photoionization from HII regions, has a 0.5-8 keV X-ray luminosity consistent with that expected from star formation. 

Using 2-10 keV luminosities and the relation given by \cite{2010ApJ...724..559L} we find that most targets have X-ray luminosities greater than would be expected from star formation ($1-9\sigma$ higher), with the exception of RGG 32, which has a hard X-ray luminosity consistent with the estimated contribution from HMXBs, and RGG B, which is not detected in the 2-10 keV band (see right panel of Figure~\ref{lx_v_sfr}). We stress again that the SFRs used are conservative upper limits, and that the actual star formation rates are lower for objects falling in the composite or AGN region of the BPT diagram.

With the exception of RGG B (and RGG 32 in the hard band), the X-ray luminosities of our targets are higher than would be expected from star formation. For our seven targets with with broad \textit{and} narrow emission line signatures of an AGN, we consider an AGN the most likely source of the nuclear X-ray emission. RGG B, which falls in the star forming region of the BPT diagram, has the lowest X-ray luminosity, and moreover, a 0.5-8 keV X-ray luminosity consistent with that which would be expected from star formation. Optical spectroscopy of RGG B was also analyzed in \cite{2016arXiv160505731B}, which carried out multi-epoch spectroscopy of star-forming dwarf galaxies with broad H$\alpha$. They found that the broad H$\alpha$ faded in most, indicating the broad emission was likely produced by a transient stellar process for those targets. RGG B was classified as ambiguous with respect to the presence of broad H$\alpha$ in spectra taken several years following the SDSS spectrum. However, even if the broad emission in the SDSS spectrum of RGG B is produced by an AGN, we cannot say with certainty that the observed nuclear X-ray emission is not associated with a HMXB.

We emphasize that more than a third of our sample falls in the composite region of the BPT diagram. Though BPT star forming objects with broad H$\alpha$ are not necessarily AGN, this work provides strong evidence to suggest that the composite objects do indeed host AGN.

\subsection{Comparison to more massive AGN}

Below, we compare the properties of the AGN considered in this paper to the properties of more massive quasars, as well as to the properties of the \cite{:kx} and \cite{2012ApJ...761...73D} samples of low-mass AGN (i.e. $M_{BH}\approx 10^{5}-10^{6} M_{\odot}$).

 \cite{2007ApJ...665.1004J} explore the ultraviolet and X-ray properties of luminous quasars in SDSS from $z=1.5-4.5$. They find a mean photon index of $\Gamma=1.92^{+0.09}_{-0.08}$, with measured values ranging from $\Gamma=1.3 - 2.3$. Our measured values of $\Gamma$ (2.25 and 2.18 for RGG 119 and 127, respectively) are consistent with these. Moreover, it is predicted that objects accreting at higher Eddington fractions will have lower (i.e. more negative) photon indices \citep{2013MNRAS.433.2485B}. \cite{2007ApJ...665.1004J} also measure $\alpha_{\rm OX}$ values ranging from -2.2 to -1.5, and find a tight relation between $l_{2500}$ and $\alpha_{\rm OX}$. While our measured $\alpha_{\rm OX}$ also fall in this range, based on the $l_{2500} - \alpha_{\rm OX}$ relation, we expect values closer to -1.0. We note that our two brightest, highest $\rm L_{bol}/L_{Edd}$ targets (RGG 119 and 127) have $\alpha_{\rm OX}\approx-1$, consistent with the \cite{2007ApJ...665.1004J} relation. 

As discussed at length in \cite{2012ApJ...761...73D}, there are several factors which can influence $\alpha_{\rm OX}$. In the disk, UV photons are inverse-Compton scattered into an X-ray corona. The temperature of the disk therefore influences how much disk energy is reprocessed in this manner. If the disk extends to the innermost stable circular orbit, the peak disk temperature is dependent on BH mass such that smaller BHs have hotter disks. The disk temperature can also be dependent on the accretion rate. Additionally, the structure of the disk, i.e., whether the disk is thin or slim, can influence $\alpha_{\rm OX}$. Finally, absorption can lead to suppressed UV emission and a harder overall spectrum. 

\cite{2012ApJ...761...73D} combine their sample of low-mass AGN with more massive AGN in order to explore trends between $\alpha_{\rm OX}$, $L_{\rm bol}/L_{\rm Edd}$, and $M_{\rm BH}$. They find $\alpha_{\rm OX}$ is not correlated at all with $L_{\rm bol}/L_{\rm Edd}$. However, they find a potential trend between $\alpha_{\rm OX}$ and $M_{\rm BH}$ such that the mean $\alpha_{\rm OX}$ decreases (becomes more negative) with increasing $M_{\rm BH}$. This is expected to be due to the blackbody temperature of the accretion disk increasing with decreasing BH mass (for a given mass accretion rate; see \citealt{2012MNRAS.420.1848D}). The accretion disk temperature is also expected to rise with increasing mass accretion rate. The mean $\alpha_{\rm OX}$ for our targets with $M_{\rm BH}\approx10^{5-6}M_{\odot}$ does not follow the trend noted by \cite{2012ApJ...761...73D}, i.e., the mean $\alpha_{\rm OX}$ is more negative than for the objects with $M_{\rm BH} \approx10^{6-7}~M_{\odot}$ (though we reiterate that our sample size is small). Additionally, there appears to be a potential trend between $\alpha_{\rm OX}$ and $L_{\rm bol}/L_{\rm Edd}$ for our sample; the objects with the highest Eddington fractions also have the least negative $\alpha_{\rm OX}$ values (see Figure~\ref{alpha_ox}). A larger sample size would aid in determining whether the disk properties of AGN in dwarf galaxies are indeed distinct from those of more massive objects.

\cite{:kx} find that their five detected objects (out of eight total) show evidence for \textit{slim disk} accretion \citep{1988ApJ...332..646A}. This accretion mode is expected to be relevant at high Eddington fractions ($L_{\rm Bol}/L_{\rm Edd}\gtrsim0.3$). In this regime, the the disk puffs up at small radii, and accretion becomes radiatively inefficient due to photon-trapping. The result is that X-ray emission appears enhanced relative to optical/UV emission. Three AGN considered here show evidence for potential slim disk accretion. Using a bolometric correction of 10, we find that RGG 119, 123, and 127 are accreting at high fractions of their Eddington luminosities ($20\%$, $54\%$, and $10\%$, respectively). Moreover, they all have relatively flat $\alpha_{\rm OX}$ values close to -1.0; \cite{:kx} find similar values. However, the remaining targets are estimated to be accreting at lower Eddington fractions and have $\alpha_{\rm OX}$ values closer to $\sim-1.5$.

\begin{figure}
\includegraphics[scale=0.6]{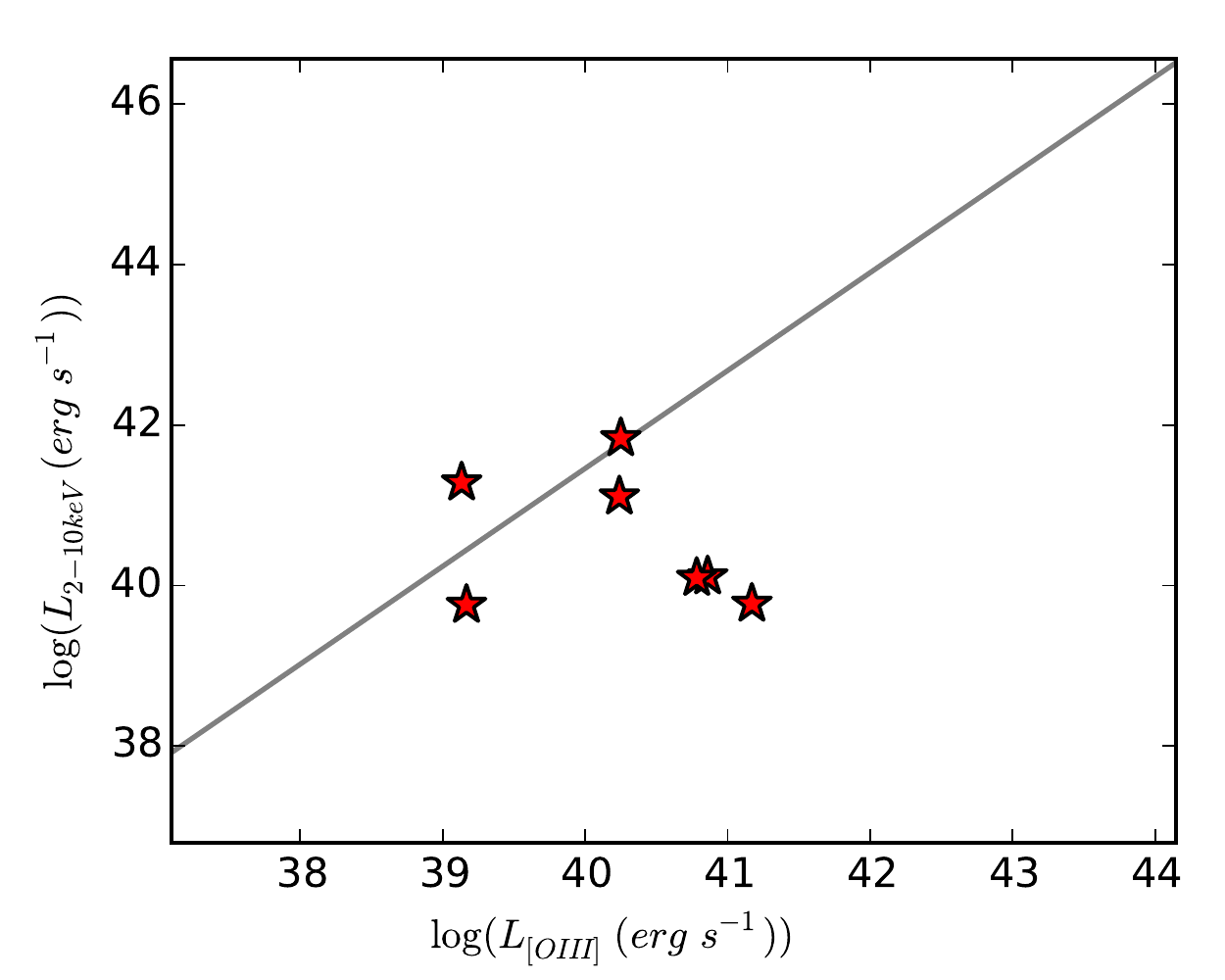}
\caption{X-ray luminosity in the 2-10 keV band versus luminosity of the [OIII]$\lambda 5007$ emission line. The gray line illustrates the relation between $L_{2-10 \rm{keV}}$ and $L_{\rm{[OIII]}}$ as defined by \cite{2006A&A...455..173P}. Our targets are shown as red stars. RGG B is not detected in the 2-10 keV band and is not included here.}
\label{xrayvoiii}
\end{figure}

Finally, \cite{2012ApJ...761...73D} explores whether their sample follow the relation between [OIII]$\lambda 5007$ luminosity and $2-10$ keV luminosity. There is a very tight correlation between these quantities for unobscured AGN \citep{2006A&A...455..173P}, since [OIII] is thought to be a good indicator of the power output of the central engine. While the \cite{2012ApJ...761...73D} objects do cluster around this relation, they find that their sample tends to scatter below it, i.e., they are relatively X-ray weak compared to the [OIII] luminosity. This behavior is similar to the Compton-thick sample from \cite{2006A&A...455..173P}, making local absorption a possible explanation for this trend. We observe this behavior for our target AGN in dwarf galaxies (Figure~\ref{xrayvoiii}). However, local absorption is not substantiated by our spectral fitting of RGG 119 and RGG 127, which finds $n_{H}$ values close to galactic values. 

\section{Conclusions}

We analyze \textit{Chandra X-ray Observatory} of 11 broad-line AGN candidates in dwarf galaxies identified in Reines et al. (2013). These include all ten objects with broad and narrow emission line AGN signatures (6 BPT AGN, 4 BPT composite), plus one low-metallicity dwarf galaxy with broad H$\alpha$ but narrow-emission lines dominated by star formation. Three out of eleven objects had \textit{Chandra} observations  analyzed in the literature. We analyze new \textit{Chandra} observations of the remaining eight, supplemented by joint HST/WFC3 F275W imaging. 

Nuclear X-ray emission is detected in all galaxies, i.e. we find a 100\% detection rate. We also find that:
\begin{itemize}
	\item{The detected X-ray nuclei are bright, with $L_{0.5-7\rm{keV}}\approx5\times10^{39} - 1\times10^{42}~\rm{erg~s^{-1}}$. Galaxies in our sample have BH masses in the range of $\sim10^{5-6}M_\odot$; we infer Eddington fractions ranging from $\sim0.1-50\%$, i.e., consistent with the range of Eddington fractions found for massive broad-line quasars. }
	\item{The observed X-ray emission in broad-line objects falling in either the AGN or composite region of the BPT diagram is brighter than would be expected from HMXBs. We conclude an AGN is the most likely source of the detected X-ray emission.}
	\item{We emphasize that the observations presented here provide strong evidence that the BPT composite objects (i.e., those thought to have contributions to the narrow-line flux from both star formation and an AGN) do indeed host actively accreting BHs.}
	\item{Our targets tend to have $\alpha_{\rm OX}$ values lower than expected based on relationships defined by classical quasars. If the measured UV emission is not significantly enhanced by nuclear star formation, AGN in dwarf galaxies seem to be X-ray weak relative to their UV emission.} 
\end{itemize}

\acknowledgments
The authors thank Marla Geha for useful suggestions. AER is grateful for the support of NASA through Hubble Fellowship grant HST-HF2-51347.001-A awarded by the Space Telescope Science Institute, which is operated by the Association of Universities for Research in Astronomy, Inc., for NASA, under contract NAS 5-26555. Support for this work was also provided by NASA through Chandra Award Number GO5-16091A issued by the Chandra X-ray Observatory Center, which is operated by the Smithsonian Astrophysical Observatory for and on behalf of the NASA under contract NAS8-03060. Support for Program number HST-GO-13943.006-A was provided by NASA through a grant from the Space Telescope Science Institute, which is operated by the Association of Universities for
Research in Astronomy, Incorporated, under NASA contract NAS5-26555.

\clearpage

\floattable
\begin{deluxetable*}{c c c c c c c c c}
\tablecaption{New \textit{Chandra} observations of broad-line AGN candidates\label{obs}}
\tablehead{
\colhead{RGG ID} & \colhead{NSAID} & \colhead{RA} & \colhead{Dec} & \colhead{BPT Class} & \colhead{Redshift} & \colhead{Obs. Date} & \colhead{Obs. ID} & \colhead{Exp. Time (ks)} }
\colnumbers
\startdata
1 & 62996 & 02:46:56.39 & -00:33:04.8 & AGN & 0.0459 & 2015-07-12 & 17032 & 17.4 \\
9 & 10779 & 09:06:13.75 & +56:10:15.5 & AGN & 0.0466 & 2014-12-26 & 17033 & 15.83  \\
11 & 125318 & 09:54:18.14 & +47:17:25.1 & AGN & 0.0327 & 2016-04-06 & 17034 & 10.26 \\
32 & 15235 & 14:40:12.70 & +02:47:43.5 & AGN & 0.0299 & 2015-04-01 & 17035 & 6.97 \\
48 & 47066 & 08:51:25.81 & +39:35:41.7 & Composite & 0.041 & 2015-12-07  & 17036 & 12.89 \\
119 & 79874 & 15:26:37.36 & +06:59:41.6 & Composite & 0.0384 & 2015-02-10 & 17037 & 10.9 \\
127 & 99052 & 16:05:31.84 & +17:48:26.1 & Composite & 0.0317 & 2015-02-22 & 17038 & 7.75 \\
B & 15952 & 08:40:29.91 & +47:07:10.4 & Star-forming & 0.0421 & 2015-01-05 & 17039 & 12.89 \\
\enddata
\tablecomments{Columns 1 and 2 give the IDs assigned by Reines et al. (2013) and the NASA-Sloan Atlas, respectively. Column 5 gives the BPT classification determined using SDSS spectroscopy. Column 7 gives the date of the \cxo observation, column 8 gives the observation ID, and column 9 gives the \cxo exposure time. }
\end{deluxetable*}

\floattable
\begin{deluxetable*}{cccccccccc}
\tabletypesize{\scriptsize}
\rotate
\tablewidth{0pt}
\tablecolumns{10}
\tablecaption{X-ray Properties\label{counts}}
\tablehead{
\colhead{RGG ID} &
\multicolumn{3}{c}{Counts} &
\multicolumn{3}{c}{Flux ($10^{-15}\rm erg~s^{-1}~cm^{-2}$)} &
\multicolumn{3}{c}{Luminosity ($\rm erg~s^{-1}$)} \\
\colhead{} & \colhead{0.5-2 keV} & \colhead{0.5-7
keV}& \colhead{2-10 keV} &\colhead{0.5-2 keV} & \colhead{0.5-7
keV}& \colhead{2-10 keV} &\colhead{0.5-2 keV} & \colhead{0.5-7
keV}& \colhead{2-10 keV}
}
\colnumbers
\startdata
1 & 2 (0.5, 5.3) & 3 (1.1, 6.7) & 1 (0.1, 3.9) &  0.66 (0.18, 1.76) & 1.42 (0.52, 3.16) & 1.15 (0.20, 4.48) & 39.52 (38.94, 39.95) & 39.85 (39.41, 40.19) & 39.76 (39.00, 40.35) \\ [0.15cm]
9 & 15 (10.3, 21.3) & 17 (12.0, 23.6) & 2 (0.5, 5.3)  & 5.23 (3.59, 7.42) & 8.57 (6.04, 11.90) & 2.52 (0.67, 6.71)  & 40.43 (40.27, 40.58) & 40.65 (40.50, 40.79) & 40.12 (39.54, 40.54) \\ [0.15cm]
11 & 2 (0.5, 5.3) & 4 (1.8, 8.0) & 2 (0.5, 5.3) & 1.37 (0.36, 3.64) & 3.97 (1.74, 7.92) & 5.01 (1.33, 13.34)  & 39.53 (38.96, 39.96) & 39.99 (39.63, 40.29) & 40.10 (39.52, 40.53) \\ [0.15cm]
32 & 8 (4.7, 13.0) & 9 (5.4, 14.2) & 1 (0.1, 3.9)  & 6.42 (3.74, 10.43) & 10.41 (6.28, 16.43) & 2.87 (0.30, 11.15) & 40.12 (39.88, 40.33) & 40.33 (40.11, 40.53) & 39.77 (38.79, 40.36) \\ [0.15cm]
48 & 48 (39.4, 58.2) & 69 (58.6, 80.9) & 21 (15.4, 28.2) & 20.60	 (16.90, 24.97) & 42.81 (36.35, 50.19) & 32.54 (23.83, 43.67) & 40.91 (40.82, 40.99) & 41.23 (41.16, 41.20) & 41.11 (40.98, 41.24) \\ [0.15cm]
119 & 457 (429.8, 485.6) & 572 (541.6, 603.9) & 117 (103.4, 132.1) & 244.7 (230.2, 260.0) & 404.10 (382.6, 426.6) & 198.1 (175.0, 223.7) &  41.93 (41.90, 41.96) &42.15 (42.13, 42.17) & 41.84 (41.79, 41.89) \\ [0.15cm]
127 & 111 (97.7, 125.8) &144 (128.9, 160.6) & 34 (26.8, 42.8) &  82.82 (72.92, 93.84) & 144.6 (129.3, 161.3) & 82.49 (64.96, 103.8) &  41.29 (41.23, 41.34) & 41.53 (41.48, 41.58) & 41.29 (41.19, 41.39) \\ [0.15cm]
B  & 2 (0.5, 5.3) & 2 (0.5, 5.3) & 0 (0, 2.3) & 0.87 (0.23, 2.30) & 1.25 (0.33, 3.32) & 0 (0, 4.1) & 39.56 (38.98, 39.98) & 39.72 (39.14, 40.15) & $<$ 40.18 \\ 
\enddata
\tablecomments{Columns 2-4 give the total counts in the 0.5-2, 0.5-7, and 2-10 keV ranges, as computed by SRCFLUX. Errors on the counts were estimated using the 90\% confidence limits given in \cite{Gehrels:1986kx}. Columns 6-8 give the corresponding fluxes, and 10-12 present luminosities.}
\end{deluxetable*}

\floattable
\begin{deluxetable*}{c c c c c c c}
\tabletypesize{\footnotesize}
\tablecolumns{7}
\tablewidth{0pt}
\tablecaption{UV properties\label{uv}}
\tablehead{
\colhead{RGG ID} & \colhead{Exp Time}& \multicolumn{3}{c}{Count rate (counts/s)} & \colhead{$f_{2500}$} & \colhead{$l_{2500}$} \\
\nocolhead{RGG ID} & \nocolhead{Exp Time} & \colhead{Total} & \colhead{Background} & \colhead{AGN} & \colhead{(${\rm erg~s^{-1}~cm^{-2}~Hz^{-1}}$)} & \colhead{(${\rm erg~s^{-1}~Hz^{-1}}$)} 
} 
\colnumbers
\startdata
1 & 747 & 1.32 & 0.37 & 0.95 & $8.18\times10^{-30}$ & $4.10\times10^{25}$  \\
9 & 900 & 6.31 & 3.19 & 3.12  & $2.69\times10^{-29}$ & $1.39\times10^{26}$   \\
11 & 882 & 32.86 & 12.59 & 20.27  & $1.75\times10^{-28}$ & $4.35\times10^{26}$   \\
32 & 747 & 37.40 & 20.33 & 17.07  & $1.47\times10^{-28}$ & $3.06\times10^{26}$   \\
48 & 801 & 5.50 & 2.66 & 2.84 & $2.44\times10^{-29}$ & $9.71\times10^{25}$  \\
119 & 747 & 29.79 & 7.63 & 22.17 & $1.91\times10^{-28}$ & $6.62\times10^{26}$   \\
127 & 756 & 5.07 & 1.38 & 3.69  & $3.17\times10^{-29}$ & $7.44\times10^{25}$  \\
B & 789 & 90.14 & 35.20 & 54.94  & $4.73\times10^{-28}$ & $1.99\times10^{27}$  \\
\enddata
\tablecomments{Column 2 gives the \textit{HST} exposure time. Columns 3-5 give \textit{HST} count rates. The total count rate refers to the total count rate in an aperture with radius of 3$''$ centered on the UV source. The background count rate refers to the count rate in the annulus immediately surrounding the inner aperture, and is intended to account for any extended star formation in the galaxy. The AGN count rate is the total count rate minus the background count rate. Columns 6 and 7 give the flux and luminosity densities at 2500$\rm{\AA}$, respectively.} 
\end{deluxetable*}

\floattable
\begin{deluxetable}{c c c c}
\tablecaption{Accretion properties\label{accretion}}
\tablehead{
\colhead{RGG ID} & \colhead{$\log(M_{BH}/M_{\odot})$} & \colhead{$L_{\rm{Bol}}/L_{\rm{Edd}}$ } & $\alpha_{\rm{OX}}$}
\colnumbers
\startdata
1 & 5.7 & 0.001 & $ -1.50^{+.16}_{-.22}$  \\
9 & 5.6\tablenotemark{a} & 0.003 & $ -1.36^{+.06}_{-.06}$  \\
11 & 4.9 & 0.013 & $ -1.89^{+.16}_{-.22}$ \\
20 & 6.1 & 0.070 & $-1.3\tablenotemark{b}$ \\
21 & 5.0\tablenotemark{c} & 0.007 & -- \\
32 & 5.2 & 0.003 & $-1.60^{+.08}_{-.09}$  \\
48 & 5.4 & 0.043 & $ -1.11\pm0.03$  \\
119 & 5.5\tablenotemark{a} & 0.197 & $ -1.08\pm0.01$  \\
123 & 5.1 &  0.543 & $-1.22\tablenotemark{b}$ \\
127 & 5.2 & 0.102 & $ -0.92\pm0.02$ \\
B & 6.1 & -- & $ -2.13^{+.16}_{-.22}$ \\
\enddata
\tablecomments{Column 2 gives BH mass estimates using the method outlined in Section 3.1. Column 3 gives the Eddington ratio, using a bolometric correction of 10. Column 4 gives the UV-to-X-ray flux ratio diagnostic. } 
\tablenotetext{a}{Black hole mass from \cite{2016arXiv160505731B}, calculated from the average of several single-epoch spectroscopic black hole mass measurements.}
\tablenotetext{b}{$\alpha_{\rm{OX}}$ measurement from \cite{2012ApJ...761...73D}.}
\tablenotetext{c}{RGG 21 (NGC 4395) has several BH mass estimates reported in the literature. For consistency, we use the value quoted in Reines et al. (2013) based on the broad H$\alpha$ emission, but note that the most recent estimate comes from \cite{2015ApJ...809..101D}, which finds a mass of $4^{+8}_{-3}\times10^{5}M_{\odot}$ based on gas dynamical modeling.}
\end{deluxetable}

\clearpage

\bibliographystyle{apj}

\end{document}